\documentclass{aa}
\usepackage{graphicx}
\begin{document}

   \title{The GSF Instability and  Turbulence do not Account for the Relatively Low Rotation Rate of Pulsars }

\author{Raphael Hirschi$^{1,2}$, Andr\'e Maeder$^3$}

     \institute{$^1$ Astrophysics Group, EPSAM Institute, University of Keele, Keele, ST5 5BG, UK. email:  r.hirschi@epsam.keele.ac.uk\\
     $^2$ Institute for the Physics and Mathematics of the Universe, University of Tokyo, 5-1-5 Kashiwanoha, Kashiwa, 277-8583, Japan\\
     $^3$Geneva Observatory, Geneva University, CH--1290 Sauverny, Switzerland.
              email:  andre.maeder@unige.ch\\              }

   \date{Received  / Accepted }

  \abstract{}{We examine the effects of the horizontal turbulence in differentially rotating stars on the GSF instability and apply our results to pre-supernova models.}{We derive the expression for the GSF instability with account  of the thermal transport and  smoothing of the $\mu$--gradient by the  horizontal turbulence.
  We apply the new expressions in numerical models of a 20 M$_{\odot}$ star.}
  {We show that if $N^2_{\Omega} < 0$ the Rayleigh--Taylor instability cannot    be killed
  by the stabilizing thermal and $\mu$--gradients, so that  the GSF instability is 
  always there and we derive the corresponding diffusion coefficient. 
  The GSF instability grows towards the very latest stages of stellar evolution.
  Close to the deep convective zones in pre-supernova stages,
  the transport coefficient of elements and angular momentum  by the GSF instability can very locally be larger than the shear instability and even as large as the thermal diffusivity. 
However the zones over which the GSF instability is acting are extremely narrow and there is not enough time left before 
the supernova explosion for a significant mixing to occur.  Thus, even when the inhibiting effects of the
$\mu$--gradient are reduced by the horizontal turbulence, the GSF instability  remains
insignificant for the evolution.} 
  {We conclude that the GSF instability in pre-supernova stages cannot be held responsible for the relatively low rotation rate  of pulsars compared to the  predictions of rotating star models.}

 \titlerunning{The GSF instability and turbulence}   
 
 \maketitle            
 
%

\section{Introduction}

The comparison of the observed rotation rate  of pulsars and stellar models in the pre-supernova
stages indicate that most stars are losing more angular momentum than currently predicted (\cite{HegerLW00,HMMXII}). Normally, the conservation of the central
angular momentum of a presupernova
model would lead to a neutron star spinning with a period of 0.1 ms, which is about two orders of magnitude faster than the estimate for
the most rapid pulsars at birth. The question has arisen whether 
 some 
rotational instabilities may  play a role in dissipating the angular momentum. We can think  in
particular of the Golreich-Schubert-Fricke (GSF)
instability (\cite{GoldreichS67,Fricke68}), which has a negligible effect in the Main--Sequence phase and  which may
play some role in the He--burning and more advanced phases (\cite{HegerLW00}), in particular when there is a very steep $\Omega$--gradient at the edge of the central dense core. This instability is generally not accounted for in stellar modeling.   The aim of this article is to examine whether the  GSF instability is important in the pre-supernova stages, when account is given to the effect of the horizontal turbulence in rotating stars which  reduces the stabilizing effects of the   $\mu$--gradient. 

Sect. 2 recalls the basic properties of the GSF instability, Sect. 3  those  of the horizontal turbulence.
The effects of turbulence  on the GSF instability are examined in Sect. 4. Sect. 5 show the results of the numerical models. Sect. 6 gives the conclusion.

\section{The GSF Instability and Solberg--Hoiland Criterion} \label{GSF}

\subsection{Recall of basics}

A rotating star with a distribution of the specific  angular momentum $j$  decreasing outwards is subject to the 
Rayleigh--Taylor  instability: an upward displaced fluid element will have a higher $j$ than the 
ambient medium and thus it will continue to move outwards. In radiative stable media, the density stratification has a stabilizing
effect, which may   counterbalance the instability resulting from the  outwards decrease of  $j$. In this respect,  
the $\mu$--gradient resulting from nuclear evolution has a strong stabilizing effect.
The stability condition is usually expressed by the Solberg--Hoiland criterion, given in the first part of Eq.~(\ref{gsf}).

The GSF instability occurs when the heat diffusion by the fluid elements  reduces the 
stabilizing effect of the entropy stratification in the radiative layers.  
 The account of a finite viscosity  $\nu$ together with  thermal diffusivity $K$ influences the instability criteria
  (\cite{Fricke68,Acheson78}). These authors found instability for  each of the two  conditions
  \begin{eqnarray}
  \frac{\nu}{K} \, N^2_{T, \, \mathrm{ad}} +N^2_{\Omega} < 0  \quad \mathrm{or} \; \; \left|\varpi \frac{\partial \Omega^2}{\partial z}\right| > \frac{\nu}{K} \, N^2_{T, \, \mathrm{ad}}  \; ,
  \label{gsf}
  \end{eqnarray}

  \noindent
  where $N^2_{T, \, \mathrm{ad}}$ is the adiabatic thermal term of the Brunt--
V\"{a}is\"{a}l\"{a} (BV) frequency and  $N^2_{\Omega}$ the rotational  contribution 
to BV for an angular velocity $\Omega$,
\begin{eqnarray}
   N^2_{T, \, \mathrm{ad}}=\frac{g \delta}{H_P} (\nabla_{\mathrm{ad}}-\nabla)\, ,\quad 
    N^2_{\Omega} \, = 
\frac{1}{\varpi^3} \, \frac{d\left( \Omega^2 \varpi^4 \right) }{d\varpi}  \, .
\label{ineg}
  \end{eqnarray}
  
  \noindent
  The viscosity $\nu=(1/3) v \ell$ represents any source of viscosity, including turbulence. 
$\varpi$ is the distance to the rotation axis  and  $z$  
the vertical coordinate parallel to the rotation axis.
  The thermal
 diffusivity $K$ is 
  \begin{eqnarray}
  K \, = \, \frac{4 \, a \, c\, T^3}{3 \, \kappa \, \varrho^2 C_{\mathrm{P}}}.
  \label{K}
  \end{eqnarray} 
  
\noindent  where the various quantities have their usual meaning.
  
 \noindent 
  \begin{itemize}
  \item The first inequality in Eq. (\ref{gsf})
  corresponds to the convective instability predicted by the Solberg--Hoiland criterion  with 
   account for the  efficiency factor 
    $\Gamma = v \, \ell/(6 \, K)$ which takes into account the radiative losses. For  $N^2_{\Omega} < 0 $, a
 displaced fluid element experiences a centrifugal force larger than in the surrounding and further moves away. 
 The first criterion in Eq. (\ref{gsf})  expresses that 
 instability  arises if the $T$ gradient, with account for thermal and viscous diffusivities, is
 insufficient to compensate for the growth of the centrifugal force during an arbitrary small displacement. 
    \item The second inequality in Eq. (\ref{gsf}) expresses a baroclinic instability related  to the
   differential rotation in the direction $z$. 
   If a fluid element is  displaced over a length $\delta z$
   in the $z$ direction, so that $\partial \Omega/\partial z \cdot  \delta z  > 0$, the angular velocity of the fluid element is larger than the local angular velocity. The excess of centrifugal force
   on this element leads to a further displacement and thus to instability.  
     It has often been concluded from this second criterion   that only cylindrical rotation laws are stable (solid body rotation being a peculiar case). This is not correct, since viscosity is never zero. In particular the horizontal
   turbulence produces a strong horizontal viscous coupling, with a large ratio $\nu/K$,
   which does not favor the  instability due to the second condition in Eq.~ (\ref{gsf}).
   \end{itemize}
   
   Numerical simulations of the GSF instability (\cite{Korycansky91}) show that the GSF instability
   develops in the form of a finger--like vortex in the radial direction, with a growth rate 
   comparable to that of the linear theory.

   \subsection{The \boldmath{$\mu$} gradient and the GSF Instability}
      
   In the course of evolution, a $\mu$ gradient develops around the convective core (there
   the $\Omega$ gradients are also large). The $\mu$ gradient produces stabilizing effects. Endal and Sofia  (\cite{EndalS78}) in their developments surprisingly use the same dependence
   on the $\mu$--gradient as for the meridional circulation (see also \cite{HegerLW00}). They apply a velocity of the GSF instability in the equatorial plane given by
   \begin{eqnarray}
   v_{\mathrm{GSF}}=
   2 \, \frac{H_T}{H_j} \, \frac{d \ln \Omega}{d \ln r} \; U_2(r)\; ,
   \label{vgsf}
   \end{eqnarray}
   \noindent
   where $U_2(r)$ is the radial component of the velocity of meridional circulation
   and $H_T$ and $H_j$ are respectively the scale heights of the distributions of $T$ and specific angular momentum $j$.
   
   Let us focus on the
   first criterion in Eq.~(\ref{gsf}), it  becomes in this case
    (\cite{KnoblochS83,Talon97}) 
  \begin{eqnarray}
  \frac{\nu}{K} \, N^2_{T, \, \mathrm{ad}} +  \frac{\nu}{K_{\mu}} N^2_{\mu}+ N^2_{\Omega} < 0  \; .
  \label{gsfmu}
  \end{eqnarray}
  
  \noindent
   $K_{\mu}$ is the particle diffusivity, either molecular or radiative. It is generally of the same order as the viscosity $\nu$, 
   thus the stabilizing effect of the $\mu$ gradient is not much reduced by  the diffusion of particles. 
   Thus, when there is a significant $\mu$ gradient, it generally  dominates and tend to stabilize
   the medium. This is why  the GSF instability is generally of only limited importance in regions with
   $N^2_{\Omega} < 0$ surrounding the stellar cores in advanced phases.
   The occurrence of horizontal turbulence, however, greatly  changes the above picture, because it is anisotropic and produces a very large particle diffusivity,
   thus reducing the effect of the $\mu$ gradient.

 \section{The Coefficient of Horizontal Turbulence in Differentially Rotating Stars}
 
 The importance of the horizontal turbulence in differentially rotating stars was  emphasized by  Zahn (\cite{Zahn92}). There are a number of observational effects  supporting its existence, in particular the thinness  of 
 the solar tachocline (\cite{SpiegelZ92}), the different efficiencies of the transport of chemical elements and of angular momentum  as well the observations of the Li abundances in solar type stars  (\cite{Chaboyer95a, Chaboyer95b}). In massive stars, 
 the horizontal turbulence increases the mixing of CNO elements
 in a favorable way with respect to observations (\cite{Mturb03}).
 
 A first estimate of the coefficient $D_{\mathrm{h}}$ of horizontal turbulence was proposed by Zahn
 (\cite{Zahn92}). A second better estimate was based on laboratory experiments with a Couette--Taylor cylinder. It gives    in a differentially rotating medium (\cite{RichardZ99,MathisPZ04}),
\begin{eqnarray}
D_{\mathrm{h}}  = \, \beta \, \varpi^3 \left|\frac{d\Omega}{d\varpi}\right| \quad \mathrm{with} \; \,\beta \approx (1.5\pm 0.5) \times 10^{-5} \; .
\label{nut} 
\end{eqnarray}

\noindent
The latitudinal variations of the
angular velocity are of the form $\Omega(r,\vartheta)=\overline\Omega(r)+\widehat \Omega(r,\vartheta) \; = \,\overline\Omega(r)+
\Omega_2(r) \,\left( P_2(\vartheta)+ \frac{1}{5}\right) $. 
$\overline{\Omega}$ is the average on an isobar, while
$\Omega_2$ expresses the horizontal differential rotation (Zahn \cite{Zahn92}, \cite{MZ04}).
\begin{eqnarray}
 \frac{\Omega_2(r)}{\overline{\Omega}(r)} \, = \,\frac{1}{5} \, \frac{r}{D_{\mathrm{h}}} 
 \left[2 \, V_2(r) -  \alpha \, U_2(r) \right] \, ,
 \label{alpho2}
 \end{eqnarray}
 \noindent
 with $\alpha \,= \, \frac{1}{2} \, \frac{d \ln(r^2 \overline {\Omega})}{d \ln r}$.
In a star with shellular rotation, one has $\Omega_2 \ll \overline\Omega(r)$. The diffusion coefficient of  horizontal turbulence (which is also the viscosity coefficient)
becomes (\cite{MathisPZ04}),
\begin{eqnarray}
D_{\mathrm{h}} &=&  \nu_{\mathrm{h}} \,=\,\, \frac{1}{2} \, \beta \, r^2 \left|\Omega_2\right|   \label{nuhfirst} \nonumber \\ [2mm]
&=& \left(\frac{\beta}{10}\right)^{1/2}
\left(r^2 \, \overline{\Omega}\right)^{1/2} 
\,\left[\, r \left|2 \, V_2 - \alpha U_2 \right|\, \right]^{1/2}  \, ,
\label{o2}
\end{eqnarray}
\noindent
where $U_2$ and $V_2$ are the vertical and horizontal components of the velocity of meridional circulation
and $\alpha$ is the same numerical factor as in Eq. (\ref{alpho2}).

The above diffusion coefficient (Eq.~\ref{nuhfirst}) derived from laboratory experiments
is essentially the definition  of the viscosity or diffusion coefficient, if the characteristic 
timescale  of the process is equal to $1/(\beta \,\Omega_2)$, i.e.
\begin{eqnarray} 
\nu_{\mathrm{h}}  \, \approx \, \frac{\ell^2}{t_{\mathrm{diff}}}\, , \quad \mathrm{with} \; \; \;
t_{\mathrm{diff}} \, \approx  \, \frac{1}{\beta \,\Omega_2}  \; ,
\end{eqnarray}

\noindent
with $\ell \sim r$. This relation implies
 that only the degree of the differential rotation in $\vartheta$  determines the importance of  horizontal turbulence. 
However, the motions on an isobar 
 in spherical geometry
are not necessarily the same as in the
Couette--Taylor experiment of rotating cylinders, which is
only a local approximation of the horizontal shear on a tangent plane.
If  the horizontal turbulence is rather related to the
differential effects of the Coriolis force (\cite{Mturb03}) which acts horizontally, i.e. $t_{\mathrm{diff}} \, \approx\, {r}/({\Omega_2 \, V_2} )^{1/2}$, one obtains the following coefficient
\begin{eqnarray}
\nu_{\mathrm{h}} =  A \, r \bigg(r
\overline{\Omega}(r) \; V_2 \;
 \left[ 2 V_2 - \alpha U_2 \right]\bigg)^\frac{1}{3} \quad \mathrm{with}\  A\leq 0.1 \;. 
 \label{dhm}
 \end{eqnarray}
 \noindent
 This expression,  despite its difference with respect 
to Eq. (\ref{o2}),  leads to  similar numerical values for the horizontal turbulence in stellar models
(\cite{MathisPZ04}), while the original estimate (Zahn \cite{Zahn92}) leads
to a coefficient $D_{\mathrm{h}}$ smaller by four orders of a magnitude. 

The expression of $\nu_{\mathrm{h}}$ requires  that we  know the vertical and horizontal components $U_2$ and $V_2$ of the velocity of meridional circulation. If not, some approximations are given in the Appendix.

\section{The Horizontal Turbulence and the GSF Instability}

 We examine what happens to the condition (\ref{gsfmu}) or Solberg-Hoiland criterion in case of thermal diffusivity and  horizontal turbulence. For that let us start from the Brunt--V\"{a}is\"{a}l\"{a} frequency in a rotating star at colatitude $\vartheta$

\begin{eqnarray}
N^2 \, = \, N^2_{T} +N^2_{\mu}+ N^2_{\Omega} \sin \vartheta \, = \nonumber \\ [2mm]
 \frac{g \, \delta}{H_P}
 \, \left( \, \nabla_{\mathrm{int}} -\nabla + \frac{\varphi}{\delta}\nabla_{\mu} \right)+
\frac{1}{\varpi^3} \, \frac{d\left( \Omega^2 \, \varpi^4 \right) }{d\varpi} \sin \vartheta  \; . \; \;
\label{Ncomplet}
\end{eqnarray}
 \noindent
 If it is negative, the medium is unstable.
 $\nabla_{\mathrm{int}}$ is the internal gradient in a displaced fluid element, while $\nabla$ is the gradient in the ambient medium. These gradients obey to the relations (\cite{M95})
 \begin{eqnarray}
 \nabla_{\mathrm{int}} - \nabla  =  \frac{\Gamma}{\Gamma+1} 
 \left(\nabla_{\mathrm{ad}} -\nabla \right) \; \mathrm{and} \; N^2_T  =  \frac{\Gamma}{\Gamma+1}  N^2_{T, \, \mathrm{ad}} \, .\quad
 \label{G+1}
 \end{eqnarray}
 \noindent
 For a fluid element moving at velocity $v$ over a distance $\ell$, $\Gamma=Pe/6=v \ell/(6 K)$, where 
 $Pe$ is the Peclet number, i.e. the ratio of the thermal to the dynamical timescale.
 $\Gamma$ is the ratio of the energy transported to the energy lost on the way
 $\ell$.
 The horizontal turbulence adds its contribution to  the radiative heat transport
 and $\Gamma$ becomes (\cite{TZ97})
 \begin{eqnarray}
 \Gamma= \frac{v \ell}{6(K+D_{\mathrm{h}})} \;.
 \end{eqnarray}
\noindent
 The ratio $\Gamma/(\Gamma+1)$ in Eq.~(\ref{G+1}) is the fraction of the energy transported.
 
The GSF instability
problem is 2D with two different coupled geometries: the cylindrical one 
associated to the rotation with the restoring force being along $\widehat{\mathbf e}_s$ and the spherical
one where the entropy and chemical stratication restoring force is along $\widehat{\mathbf e}_r$
that explains the sin$\,\vartheta$  in Eq.~(\ref{Ncomplet}), which gives the radial component of the total
restoring force. The following formula for $N^2_{\Omega}$ in spherical geometry in the case
of a shellular rotation $\overline{\Omega}(r)$ can be obtained:
 \begin{eqnarray}
N^2_{\Omega}=2 \overline{\Omega}^2 \left( 2 + 
\frac{{\rm d\,ln}\,\overline{\Omega}}{{\rm d\,ln}\,r}\right) {\rm sin}^2\,\vartheta 
+ 4\overline{\Omega}^2\,{\rm cos}^2\,\vartheta,
\label{n2o_complete}
 \end{eqnarray}
starting with Eq.~(\ref{ineg}): $N^2_{\Omega}=\frac{1}{s^3} \frac{{\rm d}(s^4\Omega^2)}{{\rm d}s}
=\frac{1}{s^3} \vec{\nabla}(s^4\Omega^2)\cdot \widehat{\mathbf e}_s$ and 
then introducing spherical coordinates. From now on, in order to simplify the problem, 
we will focus on the equatorial plane ($\vartheta= \pi/2$), in which case we simply have:

 \begin{eqnarray}
N^2_{\Omega}=2 \overline{\Omega}^2 \left( 2 + 
\frac{{\rm d\,ln}\,\overline{\Omega}}{{\rm d\,ln}\,r}\right).
\label{n2o}
 \end{eqnarray}

 The horizontal turbulence also makes some exchanges between a moving fluid element with composition given by $\mu_{\mathrm{int}}$ and its surroundings 
 with mean molecular weight $\mu$. If $f_{\mu}$ is  the amount of $\mu$ transported expressed in fraction of the external gradient, one has
 \begin{eqnarray}
 f_{\mu}=\frac{\nabla_{\mu}-\nabla_{\mu,\mathrm{int}}}{\nabla_{\mu}} \,.
 \end{eqnarray}
 
 \noindent
 One can also write  $f_{\mu}= \Gamma_{\mu}/(\Gamma_{\mu}+1)$, where $\Gamma_{\mu}$
 is the ratio of amount of $\mu$ transported to that lost by the fluid element on its way. Thus, one 
 has 
 
\begin{eqnarray}
 \nabla_{\mu,\mathrm{int}} - \nabla_{\mu}= -  \frac{\Gamma_{\mu}}{(\Gamma_{\mu}+1)} \nabla_{\mu} \, ,
 \quad  \mathrm{with} \; \Gamma_{\mu} = \frac{v \ell}{6 D_{\mathrm{h}}} \;,
 \end{eqnarray}
 \noindent
 to be compared to the first part of Eq.~\ref{G+1}. If $N^2 < 0$, the medium is unstable, thus the instability condition at the equator becomes
 \begin{eqnarray}
\left(\frac{\Gamma}{\Gamma+1}\right) N^2_{T, \, \mathrm{ad}} + \left(\frac{\Gamma_{\mu}}{\Gamma_{\mu}+1}\right) N^2_{\mu}+N^2_{\Omega}
\, < 0 \; 
\label{rim}
\end{eqnarray}
 \noindent
 The situation is similar to  the
 effect of horizontal turbulence in the case of the shear instability (\cite{TZ97}).
 
The turbulent eddies with the largest sizes  $x=v \ell/6$ are those which give the largest contribution
 to the  vertical transport. For these eddies, the equality in (\ref{rim})
is satisfied, which gives
\begin{eqnarray}
\frac{x}{x+K+D_{\mathrm{h}}} \, N^2_{T, \, \mathrm{ad}} + \frac{x}{x+D_{\mathrm{h}}} \,  N^2_{\mu} +N^2_{\Omega} = 0 \; .
\label{xdmu}
\end{eqnarray}
 
\noindent 
The diffusion coefficient by the GSF instability is $D_{\mathrm{GSF}}=(1/3) v \ell=2x$,  obtained from the solution of this second order equation, which may also be written,
\begin{eqnarray}
\left(N^2_{\mathrm{ad}}+N^2_{\mu}+N^2_{\Omega}\right)\, x^2+ \nonumber \\
\left(N^2_{\mathrm{ad}}D_{\mathrm{h}}+N^2_{\mu}(K+D_{\mathrm{h}})+N^2_{\mathrm{\Omega}}(K+2 D_{\mathrm{h}})\right) \, x+ \nonumber \\
N^2_{\Omega}(D_{\mathrm{h}}K+
D^2_{\mathrm{h}})=0 \; .
\label{2ndeq}
\end{eqnarray}

\begin{figure}
\resizebox{\hsize}{!}{\includegraphics[angle=0]{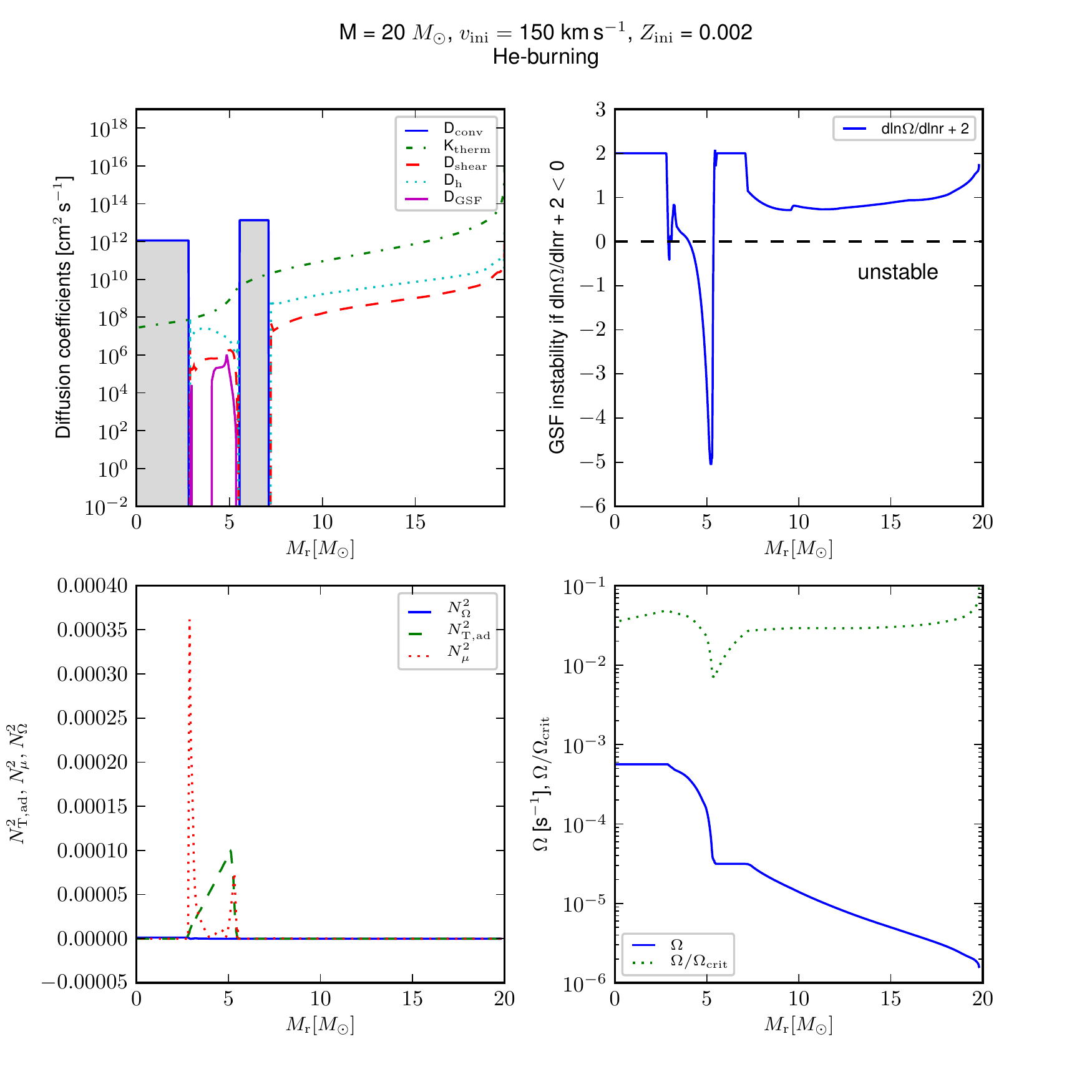}}
\caption{Properties of a 20 M$_{\odot}$ model with $Z=0.002$ and an initial rotation velocity of 150 km s$^{-1}$ during the He-burning phase, when the central
He content is $Y_{\mathrm{c}}=0.543$ and  the actual mass 19.795 M$_{\odot}$. 
a) The top left panel illustrates the various diffusion coefficients as functions of the internal mass. 
 The grey areas correspond to convective zones.
b) The top right panel 
shows the profile of the angular velocity $\Omega$--gradient $({\rm d\,ln} \,\Omega/ {\rm d\,ln} \,r)  +2$. A negative
value of this term means instability. c) The left bottom panel shows the various $N^2$. 
d) The right bottom panel shows the profile of $\Omega$  and its ratio to the local critical  
angular velocity $\Omega_{\mathrm{crit}}.$}
\label{MHe}
\end{figure}

We notice several interesting properties. 
\begin{enumerate}
\item If $N^2_{\Omega}<0$, from Eq.~(\ref{xdmu}) we see that  the GSF
instability is present in a radiative medium \emph{whatever the $\mu$-- and $T$--gradients are}. 
Thus, these  gradients cannot kill the turbulent transport by
the GSF instability. However, the size of the effects has to be determined 
for any given conditions.
\item If the diffusion coefficient  $D_{\mathrm{GSF}}$  by the GSF instability is small with respect to $K$ and $D_{\mathrm{h}}$, we have
\begin{eqnarray}
D_{\mathrm{GSF}}= 2 \frac{(-N^2_{\Omega})}{\left(\frac{N^2_{T, \, \mathrm{ad}}}{(K+
D_{\mathrm{h}})}+\frac{N^2_{\mu}}{D_{\mathrm{h}}}\right)} \; .
\label{sol1}
\end{eqnarray}

\noindent
The assumptions $D_{\mathrm{GSF}} \ll K$ and $D_{\mathrm{GSF}} \ll D_{\mathrm{h}}$
are likely, at least  at the beginning of  the GSF instability when $N^2_{\Omega}$ starts becoming negative. Nevertheless, these assumptions need
 to be verified for the cases of interest in the advanced  stages.

 \item If $N^2_{\mu}\gg N^2_{T, \, \mathrm{ad}}$, as is the case in regions surrounding
stellar cores, we get from Eq.~(\ref{xdmu})
\begin{eqnarray}
\frac{x}{x+D_{\mathrm{h}}} N^2_{\mu}+N^2_{\Omega} \approx 0 \;,
\end{eqnarray}

\begin{eqnarray}
D_{\mathrm{GSF}}\approx  2 D_{\mathrm{h}} \frac{(-N^2_{\Omega})}{N^2_{\Omega}+N^2_{\mu}}
\; .
\label{sol2}
\end{eqnarray}
\noindent
No assumption on the size of $D_{\mathrm{GSF}}$ is made here.  Due to the fast central rotation, $D_{\mathrm{h}}$ and the  $\Omega$--gradient in regions close to the central core may be  large, thus possibly favoring a significant $D_{\mathrm{GSF}}$.

\end{enumerate}

\begin{figure}
\resizebox{\hsize}{!}{\includegraphics[angle=0]{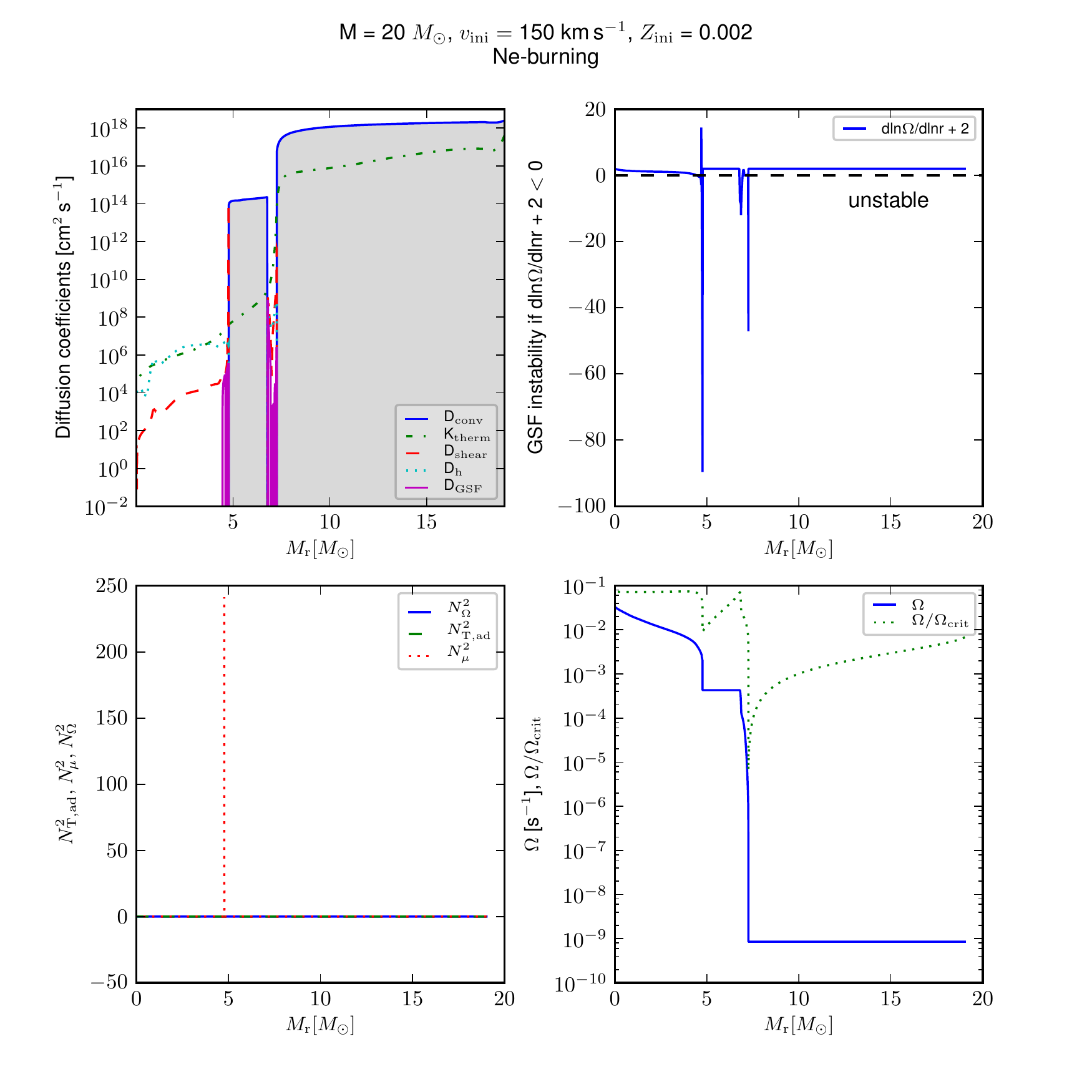}}
\caption{Same as for Fig. 1 during the phase of central Ne--burning. The actual mass is 19.412 M$_{\odot}$, the central Ne content is $X(^{20}$Ne)=0.261, the central
O--content is $X(^{16}$O)=0.726.}
\label{MNe}
\end{figure}

For more general cases, the simple solution of the second order equation  (\ref{2ndeq}) has to be  used.
Most critical of course are the values
of $N^2_{\Omega}$ and $N^2_{\mu}$, which take large values 
in a narrow region surrounding the central core in the helium and more advanced
evolutionary stages.

\section{Rotating Stellar Models in the Pre-Supernova Stages}

In order to examine quantitatively the importance of the GSF instability, we calculate the  
evolution all the way from  the Main Sequence to the Si burning stage of a 20 M$_{\odot}$ star with 
an initial rotation velocity of 150 km s$^{-1}$ with a metallicity Z=0.002  typical of the 
SMC composition. We make the choice of this composition, because the internal $\Omega$--gradients 
are steeper at lower $Z$
(\cite{MMVII}), which would favor the GSF instability. Some data for another 20 M$_{\odot}$ 
model with
an initial rotation of 300 km s$^{-1}$ are also given. Equation (\ref{2ndeq}) was used to determine the
occurence of the GSF instability and the value of $D_{\mathrm{GSF}}$.
The above expression (\ref{dhm}) for $D_{\mathrm{h}}$ is used.
The nuclear network in the advanced phases is the same as
in previous models (\cite{HMMXII}).

Figure \ref{MHe} shows in four panels the main parameters during the first part of the phase of central
He--burning. We first notice in panel d) 
the building of a $\Omega$--gradient at the edge of the convective core with a difference of $\Omega$
by about a factor of 20. This makes ${\rm d\,ln} \,\Omega/ {\rm d\,ln} \,r +2 <0$ 
in most of the region between the edge of the convective core
at 2.9 M$_{\odot}$ and the convective H--burning shell at 5.3 M$_{\odot}$ as shown in panel b). However
$N^2_{\Omega}$ 
remains negligible with respect to $N^2_T$ and $N^2_{\mu}$. 
In order to understand why, we need to look back at Eq.~(\ref{n2o}):
$N^2_{\Omega}=2 \overline{\Omega}^2 ( 2 + 
\frac{{\rm d\,ln}\,\overline{\Omega}}{{\rm d\,ln}\,r})$.
The value of 
$\Omega^2$ in the star is too small to allow a significant
value of $N^2_{\Omega}$. 
This means in fact that the centrifugal force 
in the deep interior is not strong enough to overcome the stabilizing 
effects of $N^2_T$ and $N^2_{\mu}$ as shown in panel c). 
The consequence as illustrated in panel a) is that $D_{\mathrm{GSF}}$ 
remains everywhere smaller than $D_{\mathrm{shear}}$ and is thus not significant.
We also notice that $D_{\mathrm{GSF}}$ is always much smaller
than $D_{\mathrm{h}}$ and $K$, which permits here the approximation (\ref{sol1}) made above.

\begin{figure}
\resizebox{\hsize}{!}{\includegraphics[angle=0]{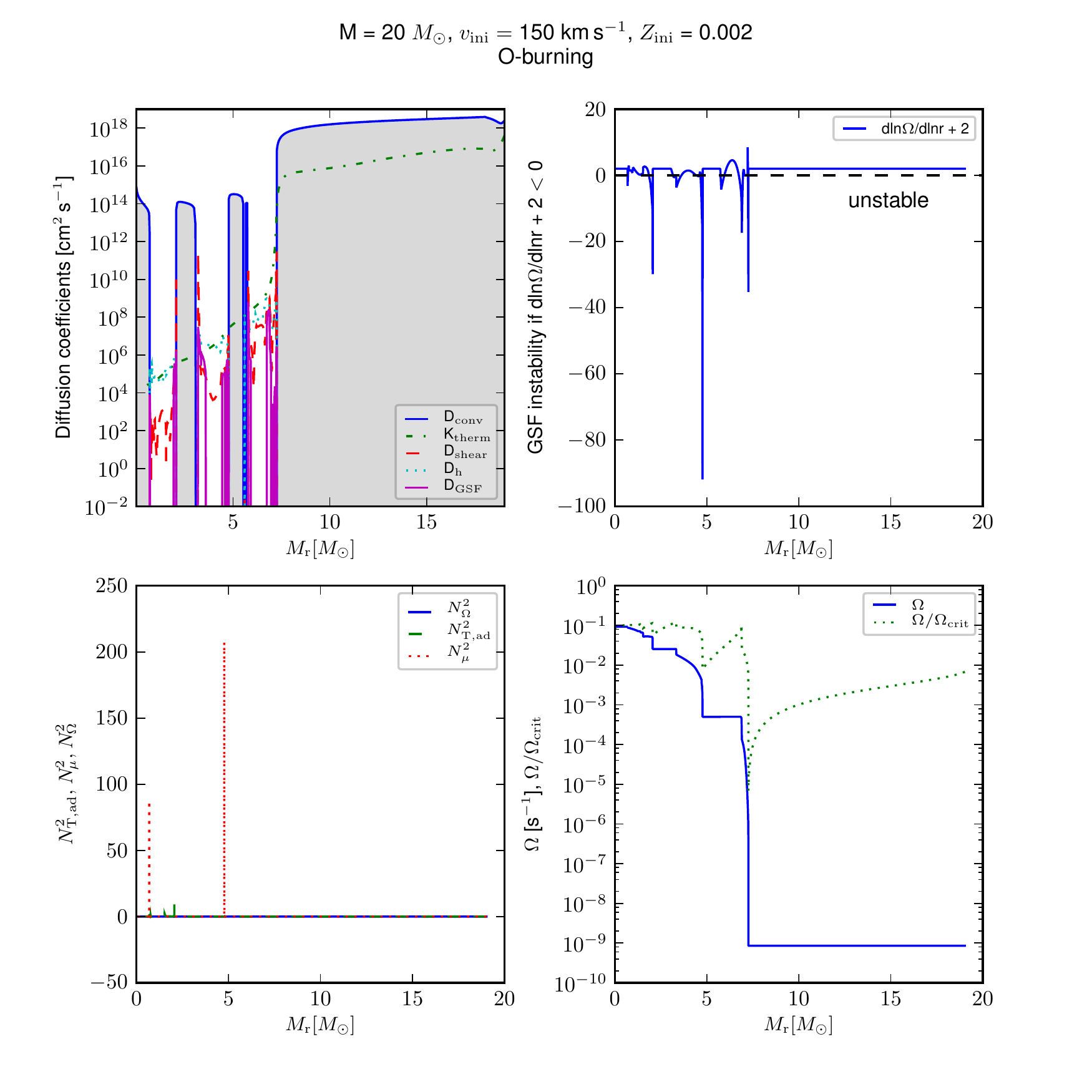}}
\caption{Same as for Fig. 1 during the phase of central O--burning. The actual mass is no longer changing, the central O--content is $X(^{16}$O)=0.692, the central
Si--content is $X(^{28}$Si)=0.153.}
\label{MO}
\end{figure}

Figure \ref{MNe} shows the same plots during the stage of central neon burning. We notice an impressive
increase 
of the central angular velocity and a very small $\Omega$
in the envelope, with a difference by a factor of $10^8$ between the two, justifying
the examination of the GSF instability. The are two "$\Omega$--walls", the big one
at  7.2 M$_{\odot}$ corresponds to  the basis of the H--rich envelope, the other one at 4.8 M$_{\odot}$ lies at the basis of the He--burning shell. The values of 
$({\rm d\,ln}\,\Omega/ {\rm d\,ln} \,r)+2$ become much more negative, however  over areas of very limited extensions.
Again, the value of $N^2_{\Omega}$ are negligible, in particular compared to the big peak of $N^2_{\mu}$ 
at 4.8 M$_{\odot}$. The result is that $D_{\mathrm{GSF}}$
is always smaller than $D_{\mathrm{shear}}$, even if very locally it can reach about the same value. $D_{\mathrm{GSF}}$ is always at least two or three orders of a magnitude smaller than $D_{\mathrm{h}}$
and $K$, permitting here the simplification (\ref{sol1}).

Figure \ref{MO} shows the situation in the central O-burning stage slightly less than a year before the central core 
collapse. Two other small steps in $\Omega$ have appeared near the center, due to the successive "onion skins" of the pre-supernova model.
We notice some new facts.  In line with what was already seen for neon burning, 
the term $ ({\rm d\,ln} \,\Omega/ {\rm d\,ln} \,r) +2$ becomes negative only in extremely narrow regions where the GSF instability
is acting with a diffusion coefficient $D_{\mathrm{GSF}}$ larger than in the previous evolutionary stages. Very locally at the upper and/or 
lower edges of intermediate convective zones,  $D_{\mathrm{GSF}}$ may even become  larger than 
$D_{\mathrm{h}}$ and $K$ reaching values above $10^8$ cm$^2$ s$^{-1}$ (there  approximation
(\ref{sol1}) is not valid!). With less than a year left before explosion,
the distance over which a significant spread may occur is about $10^{-3}$ R$_{\odot}$. 
This is not entirely negligible in the dense central regions, however
this remains of limited importance, as shown by panels b) and d) where we notice that the 
$\Omega$--walls remain unmodified despite the locally large $D_{\mathrm{GSF}}$.

\begin{figure}
\resizebox{\hsize}{!}{\includegraphics[angle=0]{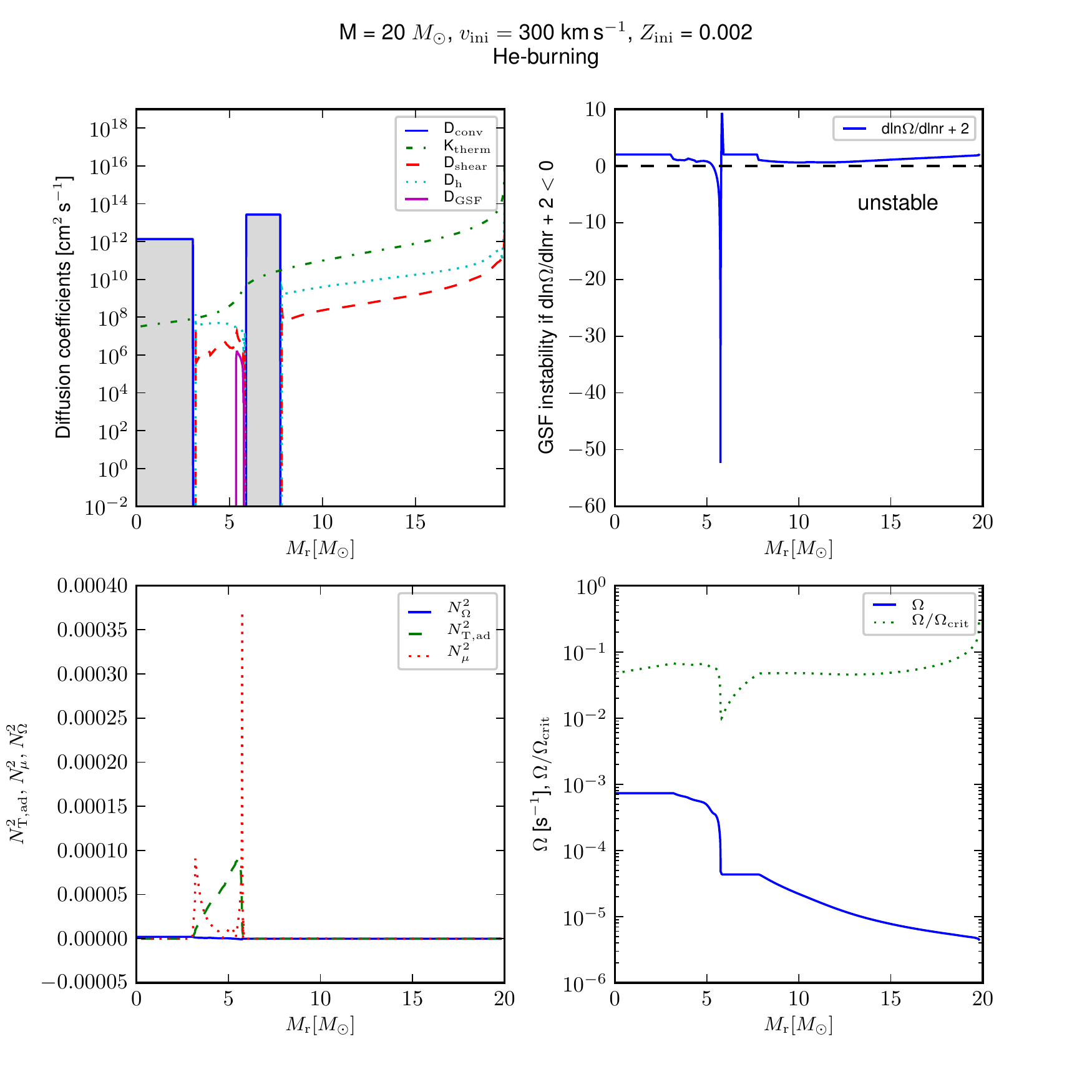}}
\caption{Same as for Fig. 1 for an initial velocity 
of 300 km s$^{-1}$. The star is in the stage of central He--burning with $Y_{\mathrm{c}}$=0.247.
 The actual mass is 19.681 M$_{\odot}$. }
\label{M404}
\end{figure}

We may wonder whether higher initial rotation velocities lead to different
results. Figure \ref{M404} shows the various panels for a similar star in the He--burning phase with an initial rotation velocity of 300 km s$^{-1}$. We see that the central rotation velocity is about the same as for the previous case of lower rotation and, in this stage which determines the further evolution,  there is no significant difference in the various properties.

\section{Conclusions}

We have examined the effects of the horizontal turbulence on the GSF instability. This instability is
present as soon as $N^2_{\Omega}$ is smaller than zero, 
whatever the effects of the stabilizing $\mu$--gradients.

On the whole, the numerical models of rotating stars show that the diffusion coefficient by the GSF instability grows towards the very latest stages of stellar evolution,
however the zones over which it is acting are extremely narrow and there is not enough time left before 
the supernova explosion for a significant mixing to occur.
Thus, even when the inhibiting effect of the
$\mu$--gradient is reduced by horizontal turbulence, the GSF instability 
is unable to  smooth the steep  $\Omega$--gradients and to significantly transport matter.

We conclude that the amplitude and spatial extension of the GSF instability makes it unable to reduce the angular momentum
of the stellar cores in the pre-supernova stages by two orders of magnitude. 
Therefore, other mechanisms such as 
magnetic fields (\cite{Sp02}, \cite{ROTMII}, \cite{MZ05}, \cite{ZBM07})
and  
gravity waves  (\cite{TC05}, \cite{MTPZ08})
must be further investigated.\\

{\bf{Appendix: some approximations} for meridional circulation}\\

The coefficient $D_{\mathrm{GSF}}$ requires, because of the horizontal turbulence, the knowledge of the components
$U_2$ and $V_2$ of the meridional circulation.
If  the solutions of the 4$^{th}$ order  system of equations governing meridional circulation are not available, some approximations may be considered. We note that the same problem would  occur for Eq.~(\ref{vgsf}) by Endal and Sofia (\cite{EndalS78}). As shown by stellar models, the orders of magnitude of $U_2$ and $V_2$ are the same. The numerical models give in general $V_2  \sim  U_2/3$ and
$\left|2V_2-\alpha U_2 \right| \sim V_2$.
 Using these orders of magnitude  in Eq.~(\ref{o2}), we get
\begin{eqnarray}
D_{\mathrm{h}}\, \approx \left(\frac{\beta}{10}\right)^{1/2}
\left(r^2 \, \overline{\Omega}\right)^{1/2}\bigg(\frac{r\,U_2}{3}\bigg)^{\frac{1}{2}}
\label{nuhapprox}
\end{eqnarray}
\noindent
For $U_2$, various expressions can be used  taking into account
the amount of differential rotation (\cite{Maeder09}). 
We can also get an order of magnitude using the approximation for a mixture of perfect gas and radiation
 with a local angular velocity $\Omega(r)$, ignoring  the effects of differential rotation on the circulation velocity  and the Gratton-\"{O}pik term which is large only  in the outer layers,
\begin{eqnarray}
  U_2(r) \, =  \, \frac{16}{9} \, \frac{\beta}{(32/3) -8 \beta -\beta^2} \,
  \frac{L(r)\; r^2}{G \, M^2_r} \nonumber \\ 
   \frac{1}{\left(\nabla_{\mathrm{ad}}-
  \nabla+ \frac{\varphi}{\delta}  \nabla_{\mu} \right)} \frac{\Omega^2 r^3}{G \,M_r}\, ,
  \end{eqnarray}
  
  \noindent
  where the various quantities have their usual meaning.

\begin{acknowledgements} We thank the referee, Dr Stephane Mathis, for his careful reading of the manuscript and his valuable comments.
R. Hirschi acknowledges support from the Marie Curie grant IIF 221145 and from 
the World Premier International Research Center 
Initiative (WPI Initiative), MEXT, Japan.
\end{acknowledgements}

{}
\end{document}